\pdfoutput=1
\documentclass[article,twocolumn,longbibliography,superscriptaddress,nofootinbib]{revtex4-2}
\usepackage{graphicx}

\usepackage{amsmath}
\usepackage{amssymb}
\usepackage{amsthm}
\usepackage{mathtools}
\usepackage{etoolbox}
\usepackage{dcolumn}  
\usepackage{multirow}

\usepackage[version=4]{mhchem}

\usepackage[noend]{algpseudocode}
\usepackage{algorithm}

\usepackage{enumerate}

\usepackage{hyperref}
\usepackage[svgnames, usenames,dvipsnames]{xcolor}
\hypersetup{hidelinks,colorlinks=true,allcolors=DarkBlue}

\usepackage{braket}

\newtheorem{proposition}{Proposition}
\AfterEndEnvironment{proposition}{\noindent\ignorespaces}

\newcommand{\Hhat}{\ensuremath{\hat{H}}}
\newcommand{\Hlochat}{\ensuremath{\hat{H}_{\rm loc}}}

\newcommand{\Ih}{\ensuremath{\hat{I}}}
\newcommand{\Xh}{\ensuremath{\hat{X}}}

\newcommand{\Zh}{\ensuremath{\hat{Z}}}

\newcommand{\Sh}{\ensuremath{\hat{S}}}
\newcommand{\Nsym}{\ensuremath{N_{\rm sym}}}

\newcommand{\psit}{\ensuremath{\psi_{\theta}}}
\newcommand{\psik}{\ensuremath{\ket{\psi}}}
\newcommand{\psikt}{\ensuremath{\ket{\psi_{\theta}}}}

\newcommand{\xvec}{\ensuremath{\mathbf{x}}}
\newcommand{\svec}{\ensuremath{\mathbf{s}}}

\newcommand{\Ns}{\ensuremath{N_{\rm s}}}
\newcommand{\Nunq}{\ensuremath{N_{\rm unq}}}
\newcommand{\nin}{\ensuremath{n_{\rm in}}}
\newcommand{\nL}{\ensuremath{n_{\rm L}}}
\newcommand{\nR}{\ensuremath{n_{\rm R}}}
\newcommand{\Ne}{\ensuremath{n_{\rm e}}}

\newcommand{\DU}{\textsc{Discard-Unphysical}}
\newcommand{\DUS}{\textsc{DU}}

\newcommand{\MU}{\textsc{Mask-Unphysical}}
\newcommand{\MUS}{\textsc{MU}}

\renewcommand{\Re}{\operatorname{Re}}
\renewcommand{\Im}{\operatorname{Im}}

\begin{document}
	\title{Autoregressive neural quantum states with quantum number symmetries}
	\author{Aleksei Malyshev}	
	\email{aleksei.malyshev@physics.ox.ac.uk}
        \affiliation{University of Oxford, Clarendon Laboratory, Parks Road, Oxford,  OX1 3PU, UK}
        \affiliation{Xanadu, Toronto, ON, M5G 2C8, Canada}
        
        \author{Juan Miguel \surname{Arrazola}}	
	\affiliation{Xanadu, Toronto, ON, M5G 2C8, Canada}

	\author{A.\ I.\ Lvovsky}
	\affiliation{University of Oxford, Clarendon Laboratory, Parks Road, Oxford,  OX1 3PU, UK}	

\begin{abstract}
Neural quantum states have established themselves as a powerful and versatile family of ansatzes for variational Monte Carlo simulations of quantum many-body systems. 
Of particular prominence are autoregressive neural quantum states (ANQS), which enjoy the expressibility of deep neural networks, and are equipped with a procedure for fast and unbiased sampling. 
Yet, the non-selective nature of autoregressive sampling makes incorporating quantum number symmetries challenging.
In this work, we develop a general framework to make the autoregressive sampling compliant with an arbitrary number of quantum number symmetries.
We showcase its advantages by running electronic structure calculations for a range of molecules with multiple symmetries of this kind.
We reach the level of accuracy reported in previous works with more than an order of magnitude speedup and achieve chemical accuracy for all studied molecules, which is a milestone unreported so far.
Combined with the existing effort to incorporate space symmetries, our approach expands the symmetry toolbox essential for any variational ansatz and brings the ANQS closer to being a competitive choice for studying challenging quantum many-body systems.
\end{abstract}
	
\maketitle

\section{Introduction}\label{sec:intro}
The variational approach has been key to numerous advances of computational quantum many-body physics, and the quest for better, more expressive and compact ansatzes is still ongoing. 
Carleo and Troyer pioneered neural networks as an ansatz for variational Monte Carlo (VMC) studies of quantum many-body systems~\cite{carleo_troyer_rbm}.
They employed a veteran neural network called restricted Boltzmann machine with complex weights and demonstrated that it could closely describe the ground states of the transverse field Ising and antiferromagnetic Heisenberg Hamiltonians. 

This result jumpstarted the field of neural quantum states (NQS). NQS were proven to be more expressive than tensor network states~\cite{nnqs_are_bigger_than_tn}, to support volume law entanglement~\cite{entanglement_in_nnqs, autoregressive_entanglement}, and to exactly represent ground states of gapped Hamiltonians~\cite{dbm_bigger_than_rbm, ite_with_dbm}.
The domain of NQS applications has been extended to quantum state tomography~\cite{nnqs_pure_tomography, nnqs_mixed_tomography}, study of open quantum systems~\cite{carleo_open_systems, autoregressive_open_systems}, and classical simulation of quantum computing~\cite{nnqs_for_quantum_computing, matija_nqs_for_qc}.
In a separate line of research, various modern neural network architectures, such as feedforward~\cite{feedforward}, convolutional~\cite{carleo_convolutional, chinese_convolutional}, recurrent~\cite{rnnqs} and transformer~\cite{transformer_j1_j2, transformer_rydberg}, were adopted as bases for NQS.

One widely exploited family of NQS is \emph{autoregressive neural quantum states (ANQS)}, introduced by Sharir {\it et~al.}~\cite{autoregressive_originals} to avoid expensive Metropolis-Hastings sampling, which was typical for early work on NQS-based optimisation.
Instead, ANQS rely on fast and unbiased \emph{autoregressive sampling}. In addition, they share the expressive power of deep neural networks, which made  them a preferred ansatz in many subsequent studies~\cite{hibbat_allah_recurrent, autoregressive_open_systems, transformer_j1_j2, transformer_rydberg, barrett_autoregressive_qchem, zhao_scalable_qchem, sharir_linear_vmc}.

Yet, existing ANQS research did not take full advantage of symmetries diagonal in the computational basis, which we refer to as quantum number symmetries.
These symmetries partition the Hilbert space of a system into non-interacting symmetry sectors and thus allow one to reduce the computational space of a problem.
For example, in the case of variational ground state search only the basis vectors from the correct symmetry sector can contribute to the ground state.

When an NQS is optimised via Metropolis-Hastings sampling, the proposal step of the latter can be easily adjusted so that irrelevant basis vectors never appear during the optimisation.
However, the standard autoregressive sampling procedure does not have a built-in proposal step.
Namely, instead of sampling from an $N$-variable probability distribution directly, one samples variables one by one (\emph{locally}), and each local probability distribution depends on the previous sampling outcomes, resulting in a tree-like sampling process. 
While samples in the correct symmetry sector can be postselected, this approach is inefficient and computationally expensive. 
It is hence appealing to  check whether a \emph{partially} sampled basis vector has a possible continuation belonging to the correct symmetry sector.
If not, one would deem it as unphysical and discard it, thus terminating the sampling early and not wasting compute power. 

In this paper, we propose an algorithm to identify unphysical partial basis vectors for a system that possesses an arbitrary number of quantum number symmetries satisfying two mild assumptions. 
This is in contrast to existing works that either relied on \emph{ad hoc} solutions to address a small number of specific symmetries~\cite{hibbat_allah_recurrent, barrett_autoregressive_qchem, zhao_scalable_qchem}, or considered quantum number symmetries of less generic nature~\cite{gauge_arnn}.
We apply our algorithm to calculations of molecular electronic structure, where we identify a class of symmetries that had previously not been considered in NQS-based quantum chemistry calculations --- specifically, the $\mathbb{Z}_2$ symmetries encoding the spatial symmetries of a molecule. 
We apply our treatment to these symmetries in addition to particle number and spin projection symmetries considered in previous works. This results in dramatically improved accuracy and computational efficiency of the variational optimisation.
In particular, we reach chemical accuracy for all studied molecules, which is a milestone unreported so far, and do this with an order of magnitude speedup compared to previous ANQS-based electronic structure calculations.

\section{Background}\label{sec:background}

\subsection{Variational Monte Carlo}\label{sec:VMC}
We consider the problem of finding the ground state of a system of $N$ interacting qubits governed by a Hamiltonian \Hhat{}.
The state of the system is described by a superposition of $2^N$ computational basis vectors $\xvec \in \lbrace0, 1\rbrace^{\otimes N}$:
\begin{equation}\label{equ:state_def}
	\psik = \sum_{\xvec = 0}^{2^N - 1} \psi(\xvec)\ket{\xvec}; \  \psi(\xvec) \in \mathbb{C}.
\end{equation}
To tame the exponential scaling of this problem we resort to variational Monte Carlo approach.
One starts by choosing an \emph{ansatz} --- a certain class of quantum states \psikt{} dependent on a parameter set~$\theta$ of polynomial size.
In this case finding the ground state can be formulated as a problem of variational minimisation:
\begin{equation}\label{equ:var_ground_state}
\arg \min_{\theta} E(\theta) =  \arg \min_{\theta}\frac{\braket{\psit|\Hhat|\psit}}{\braket{\psit|\psit}}.
\end{equation}
To solve this problem, VMC utilises the state-dependent diagonal operator of \emph{local energy}:
\begin{equation}\label{equ:local_energy_def}
     \Hlochat(\xvec) = \frac{\sum_{\xvec^\prime}\Hhat_{\xvec \xvec^\prime}\psi(\xvec^\prime)}{\psi(\xvec)},
\end{equation} 
as an unbiased estimator of the energy (we omit dependence on $\theta$ for brevity). 
That is, the energy $E(\theta)$ of a quantum state can be calculated as the expectation value of the local energy operator $\mathbb{E} \left[\Hlochat(\xvec)\right]$ taken with respect to the underlying Born distribution of the ansatz $p(\xvec) \coloneqq \frac{\left|\psi(\xvec)\right|^2}{\sum_{\xvec}\left|\psi(\xvec)\right|^2}$
~\cite{carleo_troyer_rbm}. 
In practice one evaluates this expectation after sampling a finite batch of \Ns{} basis vectors from $p(\xvec)$:
\begin{equation}\label{equ:energy_est}
E_{\rm est}(\theta) = \sum_{l=1}^{\Nunq} \Hlochat\left(\xvec^{(l)}\right) \cdot \frac{n^{(l)}}{\Ns},
\end{equation}
where we suppose that \Nunq{} unique samples were produced, and $n^{(l)}$ is the number of occurrences for the $l$-th unique basis vector.
To seek a locally optimal set of parameters, one can employ  gradient descent according to~\cite{carleo_troyer_rbm}:
\begin{multline}\label{equ:grad_def}
	\nabla_{\theta} E = 2 \Re \left\lbrace \mathbb{E} \left[  \Hlochat (\xvec) \cdot  \nabla_{\theta} \ln \psit^*(\xvec) \right] \right. \\ - \left. \mathbb{E} \left[  \Hlochat (\xvec) \right] \cdot \mathbb{E} \left[\nabla_{\theta} \ln \psit^*(\xvec) \right] \right\rbrace.
\end{multline}

\subsection{Autoregressive neural quantum states}\label{subsec:anqs}
In Ref.~\cite{carleo_troyer_rbm} Carleo and Troyer put forward neural networks as a footing for potent ansatz parameterisation thanks to their ability to compactly and accurately represent complex high-dimensional quantum states.
Many works on NQS used the celebrated Metropolis-Hastings algorithm to sample from the corresponding Born distribution. This algorithm is not without shortcomings: it produces unbiased samples from the target probability distribution only when the underlying Markov chain process has equilibrated, which in practice results in long autocorrelation times~\cite{autoregressive_originals}.
In addition, it struggles to adequately sample multimodal distributions, and in many cases the success of sampling hinges on a lucky initialisation of the underlying Markov chain.

Sharir \textit{et al.} proposed the ANQS ansatz free of these drawbacks~\cite{autoregressive_originals}.
The ANQS wave function is expressed as a product of single-qubit normalised conditional wave functions:
\begin{equation}\label{equ:anqs_def}
\begin{aligned}
	\psi(\xvec) = \prod_{i=1}^N \psi_i(x_i | \xvec_{<i}),
\end{aligned}
\end{equation}
where $\xvec_{<i} \coloneqq (x_1, x_2, \ldots, x_{i-1})$. 
The product rule~\eqref{equ:anqs_def} enables fast and unbiased sampling from the Born distribution: instead of sampling the whole basis vector at once, one can sequentially sample $N$ one-dimensional Bernoulli probability distributions $p_i(x_i|\xvec_{<i}) \coloneqq \left|\psi_i(x_i|\xvec_{<i})\right|^2$ in a tree-like manner as depicted in Fig.~\ref{fig:general_layout}A. 
One starts at the root with an empty configuration string $\xvec_{<1} = \varnothing$ and samples the  first variable $x_1$ according to the unconditional  probability distribution $p_1(x_1|\varnothing)=|\psi_1(x_1|\varnothing)|^2$, where $\psi_1(x_1|\varnothing)$ is produced by the ansatz. Based on the obtained value, one forms a partially sampled basis vector $\xvec_{<2} =x_1$ and proceeds to sampling from the respective conditional probability distribution $p_2(x_2|\xvec_{<2})$.
The process is continued until the full basis vector $\xvec = x_1x_2\ldots x_N$ is sampled. 
As a result, after a single traversal of the sampling tree one obtains an unbiased sample from $p(\xvec)$  --- this is in stark contrast with the Metropolis-Hastings sampling which might require multiple evaluations of unnormalised probability due to the possible rejections at the accept/reject step.

\subsection{Autoregressive statistics sampling}\label{subsec:stat_sampling}

As described, autoregressive sampling is already able to provide a substantial speed up over the Metropolis-Hastings sampling.
However, in the context of VMC calculations a further improvement can be achieved by \emph{autoregressive statistics sampling} proposed by Barrett {\it et al.}~\cite{barrett_autoregressive_qchem}. 
It relies on the observation that to calculate $E_{\rm est}(\theta)$ according to Eq.~\eqref{equ:energy_est}, one needs to know only the \emph{sampling statistics}, i.e. the set of pairs $\left\lbrace \left( \xvec^{(l)},n^{(l)} \right)\right \rbrace_{l=1}^{\Nunq}$.
Hence, instead of sampling a Bernoulli random variable at the tree root \Ns{} times, one might sample a single random \emph{binomial} variable $B\left(\Ns, p_1(x_1|\varnothing)\right)$ corresponding to \Ns{} trials of the underlying Bernoulli distribution  $p_1(x_1|\varnothing)$.
In practice this can be done substantially faster than sampling \Ns{} individual Bernoulli random variables.
The sampling produces two integer occurrence numbers $\nL$ and $\nR = \Ns - \nL$ indicating how many times one has to proceed to the left and to the right child trees correspondingly.
Then one applies the same trick to the second level of the sampling tree, and samples two random binomial variables $B\left(\nL, p_2(x_2|0)\right)$ and $B\left(\nR, p_2(x_2|1)\right)$ to figure out how many times each conditional distribution at the third level of the sampling tree should be sampled.
This process is repeated recursively until one reaches the $N$-th level of the sampling tree.
If for some node the sampled occurrence number is zero, it is discarded from further sampling, thus preventing exponential complexity growth.

As a result, the neural network is evaluated on only \Nunq{} configurations, as opposed to the direct sampling where the neural network is invoked \Ns{} times. 
This proved to be beneficial for the electronic structure calculations with highly peaked ground state wave functions.
In particular, Ref.~\cite{barrett_autoregressive_qchem} reports emulating sampling statistics for \Ns{} as big as $10^{12}$, which is many orders of magnitude larger than batch sizes typical for standard NQS calculations.

\section{Quantum Number Symmetries}\label{sec:qn_symmetries}

\subsection{Symmetry-aware sampling}\label{sec:symmetry_aware_sampling}
Consider a set $\lbrace \Sh_m \rbrace_{m=1}^{N_{\rm sym}}$ of operators, which commute with the Hamiltonian and are diagonal in the computational basis, i.e., $\forall m \ \Sh_m  = \sum_{\xvec}  s_m(\xvec)\ket{\xvec}\bra{\xvec}$; here $s_m(\xvec) \in \mathbb{C}$ is the  eigenvalue of $\Sh_m$ corresponding to the basis vector $\ket{\xvec{}}$. 
We refer to such set of operators as the set of \emph{Hamiltonian quantum number symmetries}, and provide a few examples --- such as the particle number and total magnetisation --- in Appendix~\ref{app:known_symmetries}. 
Since operators $\Sh_m$ are diagonal in the computational basis, they also commute pairwise, and therefore any Hamiltonian eigenstate will be an eigenstate of \emph{all} symmetry operators with the associated ordered set of eigenvalues  $\svec = \left(s_1, s_2, \ldots, s_{\Nsym} \right)$ specifying the symmetry sector of the Hamiltonian.
An important consequence is that the decomposition of the ground state $\ket{\psi_{\rm GS}}$ into the computational basis contains only the vectors with the same set of symmetry eigenvalues, which we denote $\svec_{\rm GS}$. 

It is of benefit to incorporate the latter observation into VMC optimisation. 
The simplest approach would be to compute the vector of symmetry eigenvalues $\svec(\xvec)$ for each generated sample and postselect samples in the correct symmetry sector.
However, this wastes the compute power since the samples outside the correct symmetry sector are produced too.
During the initial iterations of optimisation, the fraction of relevant samples can be extremely poor, making this approach significantly inefficient. 
Hence, we are interested in designing a sampling procedure which automatically  produces only  basis vectors from the correct symmetry sector.
We call such sampling procedures \emph{symmetry-aware}.

One can make the autoregressive sampling symmetry-aware by pruning the sampling tree on-the-fly --- that is, avoiding ``unphysical" subtrees that have no leaves in the correct symmetry sector, as illustrated in Fig.~\ref{fig:general_layout}B.
In other words, a partially sampled vector $\xvec_{<i}$ is discarded as soon as it becomes apparent that it has no physical continuations. 

Bruteforce checking all possible leaves of a given $\xvec_{<i}$ is exponentially costly. 
For some symmetries, there exist \emph{ad hoc} ways to circumvent this hurdle: e.g.~for the particle number symmetry with the target eigenvalue $n_{\rm e}$, the subtree $\xvec_{<i}$ is unphysical if $\sum_{j=1}^{i}x_j > n_{\rm e}$. Such techniques are however not readily generalisable. An additional complication is that, even if a simple rule existed for each individual symmetry, testing them one-by-one may miss a node that is unphysical because its left and right subtrees are rendered unphysical by different symmetries.
Hence it is desirable to find a physicality check algorithm that (a) has polynomial complexity with respect to $N$ and (b) checks all symmetries in combination.

\subsection{Physicality evaluation algorithm}\label{subsec:pruning_algo}
We provide such an algorithm for an arbitrary number of symmetries as long as each of them satisfies the  following two requirements:
\begin{enumerate}
	\item \emph{Local decomposability}. The eigenvalues of \Sh{} can be obtained as a composition of local eigenvalues calculated on individual qubits, i.e., $s\left( \xvec \right) = \odot_{i=1}^N s_i(x_i)$, where $\odot$ is some binary composition operation.
	For example, in the case of the particle number symmetry, $s\left( \xvec \right) = \sum_{i}x_i$, and therefore $\forall i \ s_i(x) \equiv x$ and $\odot$ is merely the addition operation.
   For symmetries with this property, we can define partial eigenvalues as $s_{<i} \coloneqq s(\xvec_{<i}) = \odot_{j=1}^i s_j(x_j)$.

	\item \emph{Polynomially-sized spectrum}. The number of unique eigenvalues of \Sh{}, either partial or not, is bounded from above by $\mathcal{O}(\textrm{poly}(N))$.
\end{enumerate}
Common quantum number symmetries --- such as those discussed in  Appendix~\ref{app:known_symmetries} --- do satisfy both of these requirements.
An important consequence of the local decomposability property is the following proposition.
\begin{proposition}
	Consider two sampling subtrees defined by the partial basis vectors $\xvec_{<i}$ and $\xvec_{<i}^\prime$. 
	If their vectors of partial eigenvalues are equal, i.e., $\svec_{<i}(\xvec_{<i}) = \svec_{<i}(\xvec_{<i}^\prime)$, then the subtrees are either both physical or not. %
\end{proposition}
\begin{proof} 
    Suppose $\xvec_{<i}$ is physical. It means that $\exists \xvec_{\geq i}: \svec({\xvec_{<i}\xvec_{\geq i}}) = \svec_{<i}(\xvec_{<i}) \odot \svec_{\geq i}(\xvec_{\geq i}) = \svec_{\rm GS}$. 
In other words, the subtree $\xvec_{<i}$ has a physical leaf ${\xvec_{<i}\xvec_{\geq i}}$.
    But then the same continuation path $\xvec_{\geq i}$ will lead to a physical leaf of $\xvec_{<i}^\prime$.
    Indeed, thanks to the local decomposability $\svec({\xvec_{<i}^\prime \xvec_{\geq i}}) =\svec_{<i}(\xvec_{<i}^\prime) \odot \svec_{\geq i}(\xvec_{\geq i})= \svec_{<i}(\xvec_{<i}) \odot \svec_{\geq i}(\xvec_{\geq i}) = \svec_{\rm GS}$.
    Hence $\xvec_{<i}^\prime$ is physical too.
\end{proof}

\begin{figure}
	\begin{algorithm}[H]
		\caption{Physicality evaluation}\label{algo:pruning}
		\begin{algorithmic}[1]
			\Statex \text{\textbf{Input:}}
			\Statex \hspace{\algorithmicindent}$i$: the index of the current sampling tree level;
			\Statex \hspace{\algorithmicindent}$\svec_{<i}$: a vector of partial eigenvalues.
			\Statex
			\Statex \text{\textbf{Output:}}
			\Statex \hspace{\algorithmicindent} \textsc{True} if $\svec_{<i}$ is physical, \textsc{False} otherwise.
			\Statex
			\Function{IsPhys}{$i;\ \svec_{<i}$}
				\If{$\left(i, \svec_{<i}\right)$ \textbf{is not in} Lookup}				
					\If{$i = N + 1$}
						\If{$\svec_{<i} = \svec_{\rm GS}$}
							\State Lookup$\left(i, \svec_{<i}\right)$ $\gets$ \textsc{True}
						\Else
							\State Lookup$\left(i, \svec_{<i}\right)$ $\gets$ \textsc{False}
						\EndIf
					\Else	
						 \State Lookup$\left(i, \svec_{<i}\right)$~$\gets$~\textsc{OR}~ $\left[\begin{array}{ll} \textsc{IsPhys}(i+1;\svec_{<i}\odot \svec_i(0)) \\ \textsc{IsPhys}\left(i+1;\svec_{<i}\odot \svec_i(1)\right).\end{array}\right.$
					\EndIf
				\EndIf

				\State \textbf{return} Lookup$\left(i, \svec_{<i}\right)$
			\EndFunction
		\end{algorithmic}
	\end{algorithm}
\end{figure}
Therefore for a given $\xvec_{<i}$ it is enough to enquire only about the physicality of the corresponding $\svec_{<i}$. 
Let us define a function $\textsc{IsPhys}(i; \svec_{<i}\coloneqq\svec(\xvec_{<i}))$ which returns \textsc{True} if $\svec_{<i}$ is physical, and \textsc{False} otherwise.
This function can be calculated recursively using the following considerations.
First, a subtree is physical as soon as at least one of its child subtrees is physical.
Second, the partial eigenvalue vectors for the left and right child subtrees are $\svec_{<i}\odot \svec_i(0)$ and $\svec_{<i}\odot \svec_i(1)$ correspondingly.
Hence, \textsc{IsPhys} satisfies the following recurrent relation: 
\begin{equation}
\textsc{IsPhys}(i; \svec_{<i}) = \textsc{OR} \left[
\begin{array}{ll}
    \textsc{IsPhys}(i+1;\svec_{<i}\odot \svec_i(0)) \\
    \textsc{IsPhys}\left(i+1;\svec_{<i}\odot \svec_i(1)\right).
\end{array}
\right.
\end{equation} 
The recursion should terminate at $i = N + 1$ with $\textsc{IsPhys}(N+1, \svec_{<N+1}) = \textsc{True}$ when $\svec_{<N+1}=\svec_{\rm GS}$ and $\textsc{False}$ otherwise. We provide the corresponding pseudocode in Algorithm~\ref{algo:pruning}.

It might seem that due to binary branching at each recursion level the runtime of the algorithm is exponential; however, one can resort to caching --- that is, store each calculated value of $\textsc{IsPhys}(\cdot)$ in a lookup table.
Since the spectra of the  symmetries considered are assumed polynomially-sized, at each level of the sampling tree we can have at most $\mathcal{O}\left(\left(\textrm{poly}(N)\right)^{N_{\rm sym}} \right) = \mathcal{O}\left(\textrm{poly}(N)\right)$ different $\svec_{<i}$, in contrast to the exponential number of subtrees.
As a result, the domain of $\textsc{IsPhys}(\cdot)$ is polynomially-sized, and so is the runtime of the algorithm.
Finally, let us note that the proposed algorithm allows targeting \emph{any} symmetry sector of the Hilbert space --- one just has to substitute $\svec_{\rm GS}$ with the desired vector of eigenvalues $\svec_{\rm ref}$ in the termination condition for \textsc{IsPhys}.

\subsection{Pruning strategies}\label{sec:local_sampling_strategies}
 \begin{figure*}
	\centering
	\includegraphics[width=\linewidth]{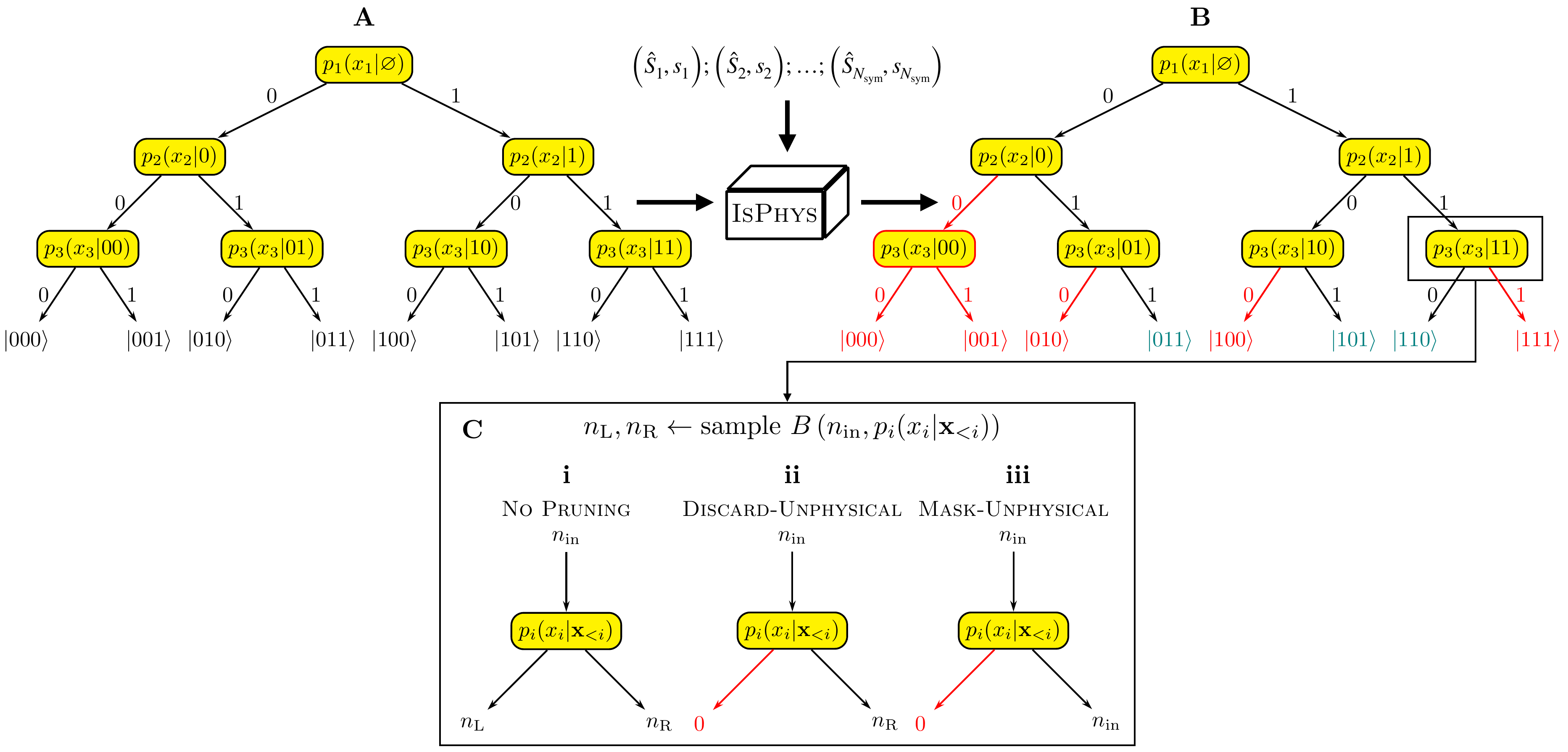}
	\caption{Symmetry-aware sampling. (A) One starts with an ordinary autoregressive sampling tree. (B) Then, one feeds the information about quantum number symmetries and their reference eigenvalues into the \textsc{IsPhys} algorithm, which identifies the unphysical subtrees (shown in red), which are pruned from further sampling. The example is for the particle number symmetry with $n_{\rm e}=2$. (C) Possible strategies for pruning unphysical subtrees.} 
	\label{fig:general_layout}
\end{figure*}

Suppose $\nin$ samples enter the node containing the local probability distribution $p_i(x_i|\xvec_{<i})$. After statistics sampling, we obtain two numbers $\nL$  and $\nR$ (with $\nL+\nR=\nin$), which should be passed to the left and right child subtrees, respectively (Fig.~\ref{fig:general_layout}C.i).
Suppose, however, that \textsc{IsPhys} shows the left child subtree to be unphysical. 
One can think of two strategies to handle this situation. 
The first approach would be to pass $\nR$ to the right subtree as planned, and pass no samples to the left subtree.
We refer to such strategy as \DU{} (\DUS, Fig.~\ref{fig:general_layout}C.ii). 
Its shortcoming is loss of samples: the occurrence numbers of final samples $n^{(l)},\ l \in \{1, \Nunq\}$ might not add up to \Ns. 

An alternative strategy was adopted in Ref.~\cite{barrett_autoregressive_qchem} and Ref.~\cite{hibbat_allah_recurrent}; we refer to it as \MU{} (\MUS{}, Fig.~\ref{fig:general_layout}C.iii). 
This strategy prescribes passing all $\nin$ ``virtual'' samples to the right subtree so that no samples are ever lost at any level of the sampling tree. 
However, this biases the sampling and the empiric probabilities $\frac{n(\xvec)}{\Ns}$ do not correspond to the probability distribution  $\left| \psi(\xvec) \right|^2$ yielded by the ansatz.
To restore the correspondence, one has to modify the conditional wave functions too: if a subtree ${\xvec_{i}0}$ of the node $\xvec_{<i}$ is unphysical, the conditional wave function must be manually modified (or \emph{masked}) so that $\psi_i^{\textsc{MU}}(0|\xvec_{<i}) = 0$ and $\psi_i^{\textsc{MU}}(1|\xvec_{<i}) = 1$ regardless of their initial values. This modified amplitudes are then used in all subsequent calculations.

Unfortunately, masking might bring unforeseen side effects affecting the ansatz expressibility.
For example, for a particle number symmetry the value of $x_N$ is completely defined by the sampled values of $\xvec_{<N}$, and therefore for \emph{all} $\xvec_{<N}$ the conditional wave function $\psi_N(x_{N}|\xvec_{<N})$ will be masked.
As a result, the subnetwork encoding $\psi_N(x_{N}|\xvec_{<N})$ is rendered useless.
A possible remedy is to apply \MUS{} to the first $N-d$ levels of the tree, and then use \DUS{} for the last $d$ levels (where $d$ is a hyperparameter).
We denote such family of strategies as \MUS-$d$.
The purpose of \MUS-$d$ strategies is to maintain a higher level of variational freedom at the cost of sample loss in a controllable way.

\section{Results}\label{sec:results}

\subsection{System description}\label{sec:system_descr}
The above discussion of symmetry-aware sampling is system agnostic: it applies to any system of interacting qubits. 
In this section, we showcase the advantages of our approach on the problem of molecular electronic structure calculation, since molecules usually possess multiple quantum number symmetries.

We work in the Born-Oppenheimer approximation and in the second quantisation approach, which means that the Hilbert space is spanned by $N$ spin-orbitals, each of which can be occupied by one of $n_{\rm e}$ electrons. 
The system wave function is then represented as a linear combination of Slater determinants (SDs) comprised of \Ne{} spin-orbitals out of $N$.
The system Hamiltonian is a conventional fermionic Hamiltonian including one- and two-body interactions:
\begin{equation}\label{equ:sq_hamiltonian}
\hat{H}_{\rm SQ} = \sum_{ij}h_{ij}\hat{a}_i^\dagger \hat{a}_j + \sum_{ijkl}h_{ijkl}\hat{a}_i^\dagger \hat{a}_j^\dagger \hat{a}_k \hat{a}_l.
\end{equation}

We obtain the basis of molecular spin-orbitals with the Hartree-Fock (HF) procedure.
In this case the mean-field  solution (the HF state) is represented as a single Slater determinant, in which  %
the first \Ne{} lowest energy orbitals are filled and the rest are empty \cite{mcardle_review}. 
We use the Jordan-Wigner encoding to map this fermionic system into qubit form. 
This preserves the occupation number interpretation, so the HF state is represented by the basis vector $\ket{\xvec_{\rm HF}} \coloneqq \ket{\underbrace{11\ldots1}_{\Ne}\underbrace{0\ldots0}_{N - \Ne}}$.

\begin{table*}
\begin{ruledtabular}
\begin{tabular}{cccccc}

\multirow{2}{*}{Molecule} & \multirow{2}{*}{$N$} & \multirow{2}{*}{$n_{\rm e}$} & \multirow{2}{*}{Number of $\mathbb{Z}_2$ symmetries} & \multicolumn{2}{c}{Computational space size (SDs)}\\
&&&& With $\mathbb{Z}_2$ &  Without $\mathbb{Z}_2$ \\
\hline
\ce{LiH}& 12 & 4 & 4 & 69 & 225\\
\ce{H2O}& 14 & 10 & 3 & 261 & 441\\
\ce{NH3}\footnotemark[1]& 16 & 10 & 2 (3) & 3,136 (1,576)& 3,136\\
\ce{CH4}\footnotemark[1]& 18 & 10 & 2 (4) & 15,876 (4,076) & 15,876\\
\ce{N2}& 20 & 14 & 5 & 1,824 & 14,400\\
\ce{C2}& 20 & 12 & 5 & 5,612 & 44,100\\
\ce{LiF}& 20 & 12 & 4 & 11,124 & 44,100\\
\ce{PH3}\footnotemark[1]& 24 & 18 & 2 (3) & 48,400 (24,202) & 48,400\\
\ce{LiCl}& 28 & 20 & 4 & 250,581 & 1,002,001\\
\ce{Li2O}\footnotemark[1]& 30 & 14 & 4 (5) & 10,355,569 (5,179,569)& 41,409,225\\
\ce{NaCl}& 36 & 28 & 4 & 2,341,648 & 9,363,600\\
\end{tabular}
\end{ruledtabular}
\footnotetext[1]{Asymmetric geometry.}
\caption{\label{tab:mol_ref_table} Reference information for studied molecules. The ``asymmetric geometry'' superscript indicates molecules which exist in geometries with larger number of spatial symmetries compared to the geometries provided by the PubChem database~\cite{pubchem} and studied in Ref.~\cite{barrett_autoregressive_qchem} and~\cite{zhao_scalable_qchem}. For such molecules we provide in parentheses figures corresponding to the most symmetric geometries.}
\end{table*}
\subsubsection*{Molecular symmetries}\label{subsubsec:mol_symmetries}
We consider three types of quantum number symmetries inherent to molecules.
Two of them --- the total number of electrons $\hat{n}_{\rm e}$ and their total spin projection $\hat{S}_z$ --- were incorporated in previous research~\cite{barrett_autoregressive_qchem, zhao_scalable_qchem} and are discussed in Appendix~\ref{app:known_symmetries}.
The third type are $\mathbb{Z}_2$ symmetries encoding spatial symmetries of a molecule; they are considered for the first time in the context of NQS-based quantum chemistry calculations. 

One usually accounts for spatial symmetries by considering irreducible representations (irreps) of the molecule point group, which is a group of spatial transformations preserving the nuclear positions.
It is known that the ground state of a molecule belongs to one of irreps of the largest abelian subgroup of the point group.
At the same time, each Slater determinant belongs to one of the irreps too, and therefore only the SDs within the same irrep as the ground state contribute to the latter.

To include this selection rule into ANQS-based calculations, we build on the observation made by Setia {\it et al.}~\cite{point_group_tapering}.
It states that after the Jordan-Wigner transform every spatial symmetry in the largest abelian subgroup of the point group would translate into a Hamiltonian $\mathbb{Z}_2$ symmetry. 
A $\mathbb{Z}_2$ symmetry is an operator $\Sh$ which commutes with \Hhat{} and is an $N$-fold tensor product of either the identity or Pauli \Zh{} matrices.
For example, if $\Hhat = \Xh_1\Xh_2 + 0.5 \Zh_1\Zh_2$, then $\Zh_1\Zh_2$ is its $\mathbb{Z}_2$ symmetry, while $\Ih_1 \Zh_2$ and $\Zh_1 \Ih_2$ are not.
All $\mathbb{Z}_2$ symmetries of a Hamiltonian could be found in an automated way using the algorithm outlined by Bravyi {\it et al.}~\cite{bravyi_tapering}. 

An important consequence is that a basis vector belongs to the ground state irrep only if for each $\mathbb{Z}_2$ symmetry it has the same eigenvalue as the ground state. 
However, at the start of calculations the ground state and its symmetry sector are unknown.
Therefore, to fix the symmetry sector we assume that the HF Slater determinant will necessarily contribute to the ground state, which is a valid premise for a vast number of molecules.
Consequently, for the purpose of running the $\textsc{IsPhys}$ algorithm, we fix $\svec_{\rm GS}$ to be equal to $\svec(\xvec_{\rm HF})$.

\subsection{Numerics}\label{subsec:numerics}

\subsubsection{General results}
\begin{figure*}
	\centering
	\includegraphics[width=0.88\linewidth]{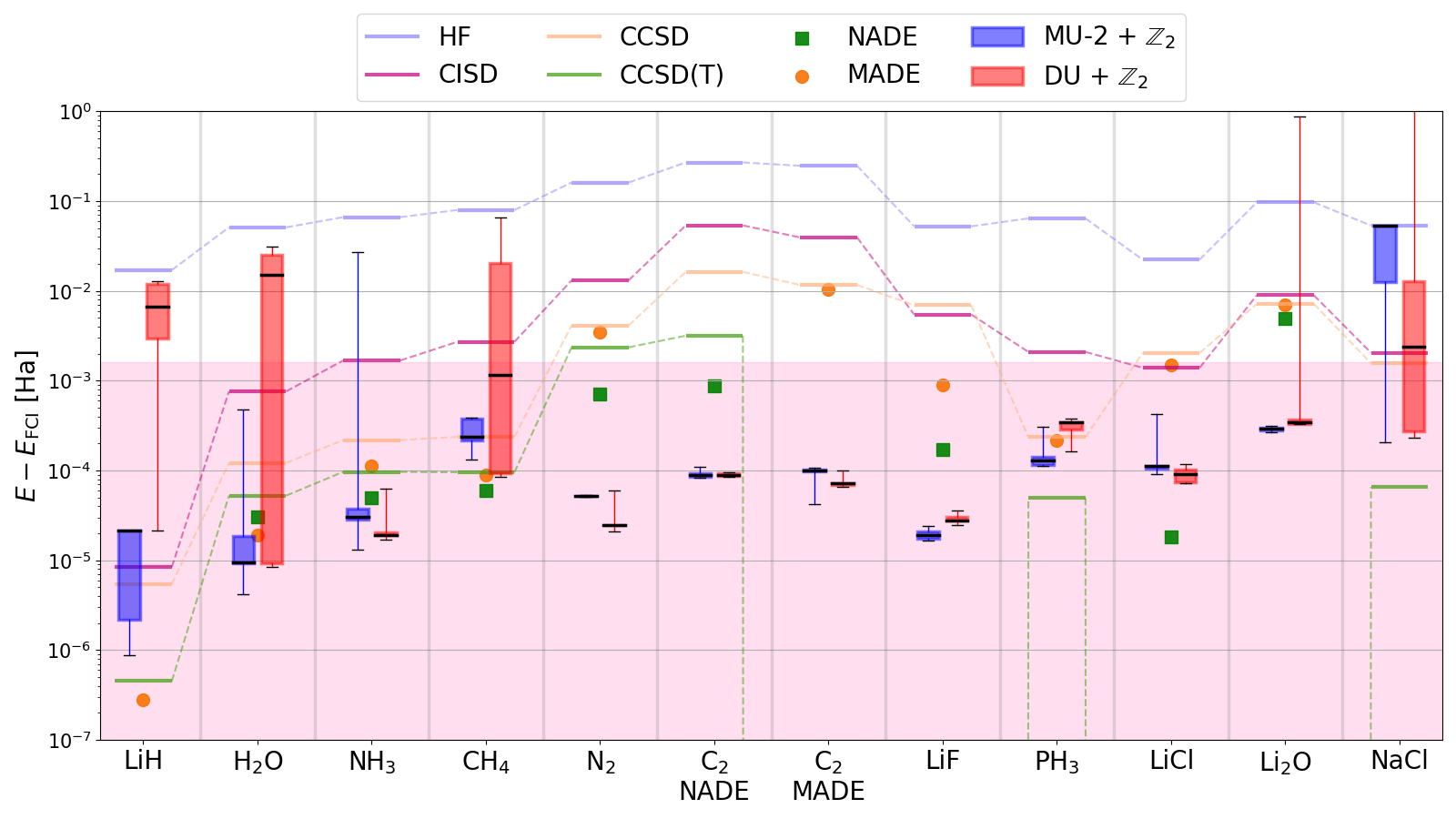}
	\caption{Comparison of the variational energies achieved by our symmetry-aware ANQS-based optimisation and the existing ANQS variants studied in the literature~\cite{barrett_autoregressive_qchem, zhao_scalable_qchem} (denoted as ``NADE'' and ``MADE'' respectively). Ref.~\cite{barrett_autoregressive_qchem} and Ref.~\cite{zhao_scalable_qchem} studied the C$_2$ molecule at two different geometries, we present results for both of them. The black bold line on the box bodies corresponds to the median value and whiskers stretch from the minimum to maximum value in the distribution of results. The shadowed area spans energies below the chemical accuracy benchmark. For better visibility we plot the reference energies of different methods as continuous curves, even though they belong to different molecules and are not related. For \ce{C2} (MADE), \ce{LiF}, \ce{LiCl} and \ce{Li2O} molecules CCSD(T) energies are below the corresponding FCI energies, and therefore the CCSD(T) curve ``dips'' due to logarithmic scale of the energy error axis.} %
	\label{fig:final_results_with_references}
\end{figure*}

In the first set of experiments we test our ANQS on a large set of molecules. The reference information for the  molecules studied, including the number of $\mathbb Z_2$ symmetries and the corresponding computational space sizes is presented in Table~\ref{tab:mol_ref_table}. The $\mathbb{Z}_2$ symmetries were calculated for each qubit Hamiltonian using the implementation of the  Bravyi {\it et al.} algorithm~\cite{bravyi_tapering} provided by the PennyLane software library~\cite{pennylane_main, arrazola2021differentiable}. For each molecule we employ five randomly initialised instances of ANQS corresponding to different seeds of the underlying pseudorandom number generator. We test both the  \MUS{}-2 and \DUS{} pruning strategies because our initial ablation studies (Appendix~\ref{app:ablation_studies}) showed their performance to be comparable and substantially surpassing that of \MUS{}-0.

Each optimisation lasts for $3\times10^4$ iterations. We vary the sample batch size $N_s$ across the simulation: to reduce the computational burden of the first iterations when many basis vectors might be sampled, we start with $\Ns = 10^5$; we increase it afterwards in a stepwise manner several times during the optimisation until it reaches the final value of $10^8$. After the gradient calculation, the ANQS parameters are updated with the ADAM optimiser in a default configuration of hyperparameters.
More details and the description of the ansatz architecture are provided in Appendix~\ref{app:experiment_details}.

The results are given in Fig.~\ref{fig:final_results_with_references}, where we measure the difference between the minimum variational energy achieved in the course of optimisation and the full configuration interaction energy $E_{\rm FCI}$, which is a quantum chemistry parlance for exact diagonalisation.
We also plot reference energy difference values obtained with such conventional quantum chemistry methods as CISD, CCSD and CCSD(T) (see Ref.~\cite{mcardle_review} for an overview).

We compare our results with the minimum energy errors obtained in the existing works~\cite{barrett_autoregressive_qchem} and~\cite{zhao_scalable_qchem} (which did not account for the $\mathbb{Z}_2$ symmetries and used the \MUS{} strategy).
Another difference is  that both of the above references aimed to keep the final \Nunq{} within some range  and thus the initial batch size $N_s$ was adaptively chosen at each iteration after multiple possible resamplings.
Their motivation was to sample not too few unique basis vectors --- in which case the gradient becomes too noisy --- but also not too many, since the local energy calculations might become prohibitively expensive.
In our experience, there is no need for adaptive batch sizes so long as \Ns{} is large enough.

For all molecules, our method with both pruning strategies demonstrates  median errors below the chemical accuracy of 1.6 mHa --- this is a milestone unreported so far.
We also observe that the largest gain in accuracy (up to an order of magnitude compared with previous research) is achieved for molecules with higher spatial symmetry such as LiCl, N$_2$, LiF, Li$_2$O and C$_2$.
We note that our method appears to underperform with respect to Barrett {\it et al.} on LiCl molecule. 
However, the experimental point shown for Ref.~\cite{barrett_autoregressive_qchem} in Fig.~\ref{fig:final_results_with_references} corresponds to the best performance amongst multiple seeds; the average energy of the Barrett {\it et al.} optimisation on LiCl is above the median value obtained in this work.
 
\subsubsection{Time to CCSD}\label{sec:time_to_ccsd}
\begin{figure}
	\centering
	\includegraphics[width=\linewidth]{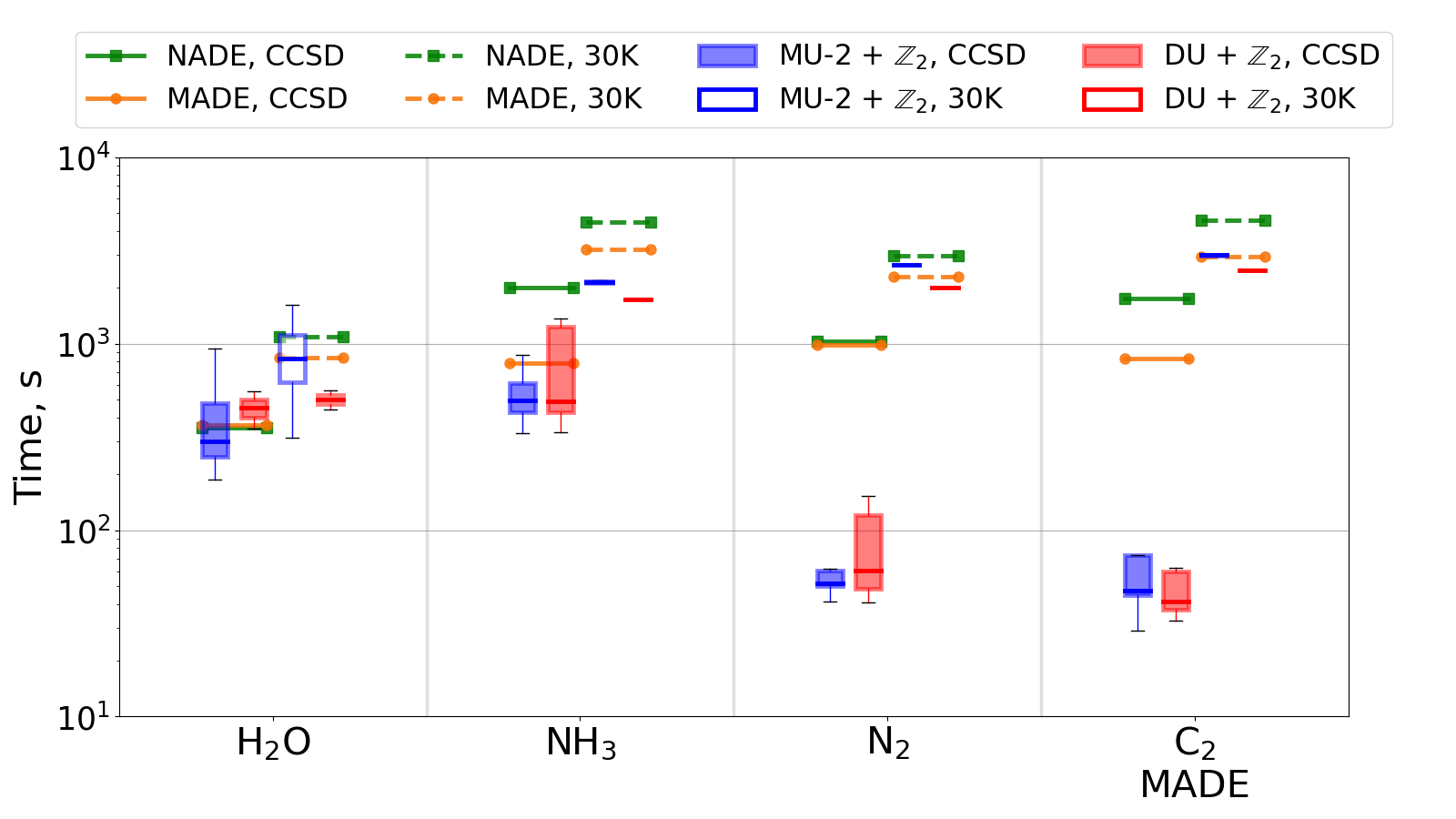}
	\caption{Comparison of the computational performance of the proposed symmetry-aware ANQS-based optimisation and the existing ANQS variants. Data points labelled ``CCSD'' correspond to the time required to achieve the CCSD level of accuracy; data points labelled ``30K'' correspond to the time spent on $3\times10^4$ variational optimisation iterations.}
	\label{fig:ccsd_timings}
\end{figure}
While the final accuracy achieved during the variational optimisation is deemed to be the primary figure of merit, it is also important to analyse the computational cost of the method.
To that end, we follow Ref.~\cite{zhao_scalable_qchem} and extract two additional metrics from the experiments held in the previous section --- the time required to achieve the CCSD level of accuracy and the total time spent on $3\times 10^4$ iterations; we present the results in Fig.~\ref{fig:ccsd_timings} together with the numbers from Ref.~\cite{zhao_scalable_qchem}.

As can be seen, the per-iteration runtime of our method is similar to that of previous approaches.
However, for highly symmetric N$_2$ and C$_2$ molecules our method converges to the desired level of accuracy much faster than the existing ones: we observe a speedup of more than an order of magnitude. 
Another example is Li$_2$O: the ANQS of Ref.~\cite{barrett_autoregressive_qchem} required 45.6 hours to achieve the accuracy of $1.8 \cdot 10^{-3}$ Ha, which is above the chemical accuracy (not shown on the plot, we quote the figure given in Ref.~\cite{barrett_autoregressive_qchem}).
In contrast, our method required 5.1 hours on average to reach the chemical accuracy.%

\subsubsection{Loss of samples}\label{sec:loss_of_samples}
\begin{figure*}
	\centering
	\includegraphics[width=0.4\linewidth]{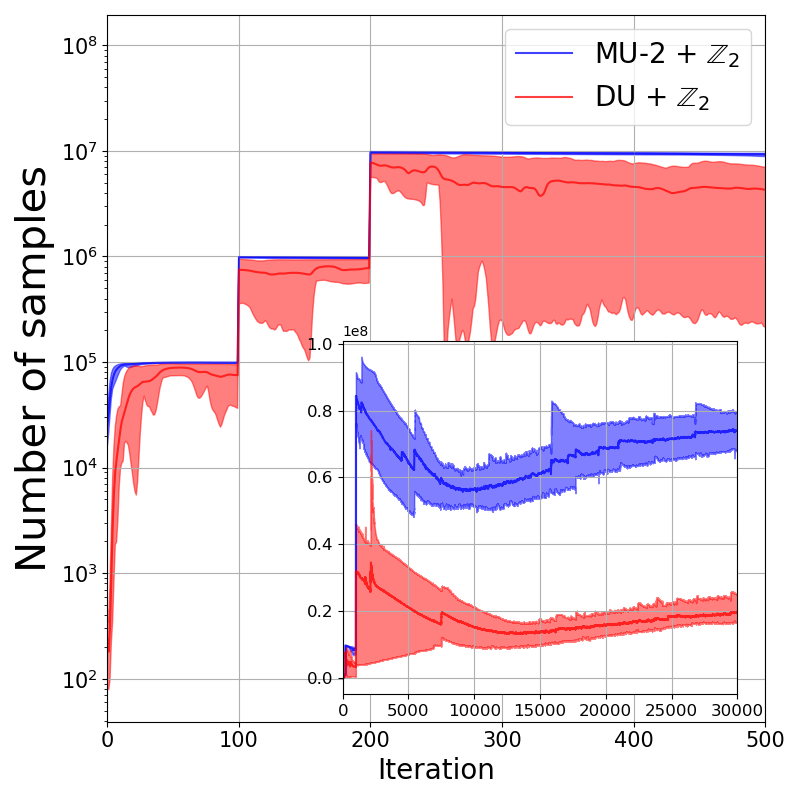}
    \includegraphics[width=0.4\linewidth]{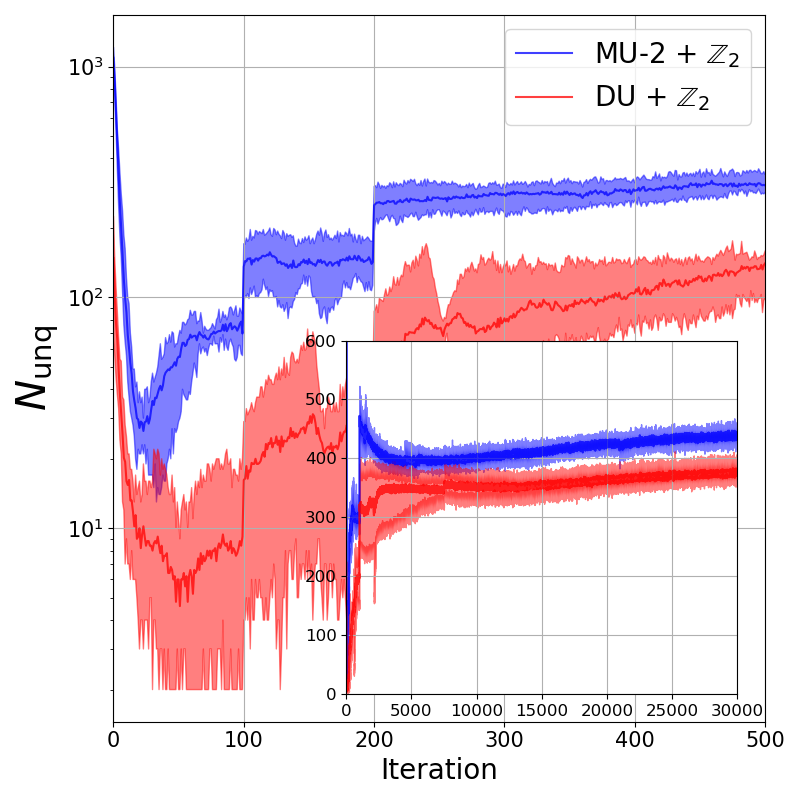}
	\caption{Comparison of the \MUS{}-2 and \DUS{} pruning strategies with respect to the total number of samples and the number of unique samples produced at every iteration. The solid lines represent the median values obtained during five runs of randomly initialised ANQS, while shaded regions span from minimum to maximum values. The main plots focus on the first 500 iterations when the variational energy passes the HF reference value. The insets show the performance of both strategies over the whole course of iteration. Overall, the \MUS{}-2 strategy loses fewer samples and produces more unique samples at each iteration. }
	\label{fig:loss_of_samples}
\end{figure*}
In the experiments described so far, both \MUS{}-2 and \DUS{} strategies performed on par.
To reveal the difference between them, we investigate how the total number of samples and the number of unique samples produced per iteration changes in the course of optimisation.

In Fig.~\ref{fig:loss_of_samples} we take \ce{N2} as a model molecule and show how typical optimisation unravels for both strategies. 
At the very start, the ANQS is equally likely to sample any physical basis vector, and therefore first iterations feature large \Nunq. 
In addition, a substantial number of samples is lost by both strategies since the probability mass assigned to the correct symmetry sector is roughly equal to the fraction of the unmasked Hilbert space it occupies.
Yet, the ansatz quickly learns the peaked structure of the molecular wave function, and as the variational energy passes the Hartree-Fock reference value (which can be achieved with only one basis vector contributing to the state), \Nunq{} reaches minimum.
At this stage, a large portion of the probability mass is located inside the correct symmetry sector, and therefore few samples are lost. 
Finally, as the optimisation proceeds, \Nunq{} gradually increases, which reflects how the ANQS seeks to decrease the energy by adding more and more basis vectors to the quantum state. 

While both strategies perform qualitatively similarly, there is a quantitative difference between them: the \DUS{} strategy is more likely to lose samples, whether unique or not --- sometimes it produces as few unique samples as one.
One might not consider this as a major drawback: neither the total number of samples nor the number of unique samples are a primary figure of merit for the variational optimisation; in the end,  \DUS{} achieves lower variational energies on \ce{N2} molecule than \MUS{}-2.
However, we see this as a disadvantage of \emph{practical} importance: for bigger molecules the \DUS{} strategy is more likely to produce no samples in the correct symmetry sector at early stages of optimisation, and thus stall the whole process.
Even though this can be mitigated by carefully scheduling \Ns{}, we believe this puts \MUS{}-2 forward as a more robust and practical pruning strategy.

\section{Conclusion}\label{sec:conclusion}
We proposed a systematic approach to include an arbitrary set of quantum number symmetries into ANQS-based optimisation.
We used it to carry  out electronic structure calculations on a set of molecules and achieved previously unattainable variational energies with an order of magnitude speedup.
In addition, we showed that $\mathbb{Z}_2$ symmetries initially conceived in the context of variational quantum algorithms are also beneficial in the context of NQS. They allow one to reduce the computational space of the problem to the level operated with by the conventional quantum chemistry methods.

Now that both space~\cite{markus_review} and quantum number symmetries are in the toolbox of ANQS-based optimisation, it would be interesting to see how much further it is possible to push the latter.
In that regard, one might follow the ideas described by Sharir {\it et al.}~\cite{sharir_linear_vmc} to reduce the computational burden of the local energy calculations, which remained the bottleneck in our experiments.

\section*{Acknowledgements}
AM thanks Davide Castaldo, Matija Medvidovi\'c, Alain Delgado Gran and Soran Jahangiri for the fruitful discussions and kind responses given promptly to any request.
\bibliography{bibliography}

\appendix

\section{Common quantum number symmetries}\label{app:known_symmetries}
Here we overview some quantum number symmetries considered in the literature. 
All of these symmetries are \emph{additive} in that the composition operation $\odot$ is just the binary addition.
We discuss  $\mathbb{Z}_2$ symmetries --- which are \emph{multiplicative} --- in Sec.~\ref{sec:system_descr} of the main text.

For first-quantised systems, such as chains of spin-$\frac12$ particles, where the computational basis vectors represent spin configurations, $\ket{0}$ can denote $\ket{\uparrow}$ and $\ket{1}$ --- $\ket{\downarrow}$. The   %
\emph{ total magnetisation}~\cite{hibbat_allah_recurrent} 
is then calculated as $s(\xvec) = \sum_{i=1}^{N}\left(\frac{1}{2} - x_i\right)$. 
    
For second-quantised systems (such as molecules studied in this paper) where the computational basis vectors bear the meaning of occupation strings, two symmetries can be identified.
\begin{itemize}
    \item \emph{Total number of particles $\hat{n}_{\rm e}$}. The eigenvalues of this symmetry are calculated as $s_{\hat{n}_{\rm e}}(\xvec) = \sum_{i=1}^{N}x_i$. Mathematically, this symmetry partitions the $N$-qubit Hilbert space into the same symmetry sectors as the total magnetisation symmetry in spin chains, albeit each sector is labelled with a different eigenvalue. 
    \item \emph{Total spin projection $\hat{S}_{z}$}~\cite{barrett_autoregressive_qchem, zhao_scalable_qchem}. In the context of electronic structure calculations qubits represent spin-orbitals. %
    The spin-orbitals are usually ordered so that the odd orbitals have spin projection $s_z = \frac{1}{2}$ and the even ones have $s_z = - \frac{1}{2}$.  
    The total spin projection of the ensemble of electrons represented by a computational basis vector is calculated as a sum of spin projections of occupied orbitals. 
    Correspondingly, the eigenvalues of this symmetry are calculated as $s_{\hat{S}_z}(\xvec) = \frac{1}{2}\sum_{i=1}^{N}(-1)^{(i-1)}x_i$.
    \end{itemize}

    Each of these symmetries in a system with $N$ qubits has at most $N+1$ different eigenvalues. 
    
\section{Experiment details}\label{app:experiment_details}
\subsection{ANQS architecture}\label{app:anqs_architecture}
The autoregressive ansatz used in our numerical experiments is comprised of $N$ subnetworks following the general NADE architecture~\cite{nade}; the $i$-th subnetwork computes $\log \psi_i(x_i|\xvec_{<i})$ (the logarithm operation is applied for numerical stability).

Each subnetwork is represented with a fully-connected multilayer perceptron (MLP) equipped with complex weights.
Each MLP takes $i - 1$ bits as input and produces two complex numbers ($\log \psi_i(0|\xvec_{<i})$ and $\log \psi_i(1|\xvec_{<i})$) as output. 
The first MLP corresponds to the unconditional wave function $\psi_1(\cdot)$ and therefore has no input. 
Each MLP has two hidden layers of width 64.
We use the following sequence of nonlinear activations:
\begin{align*}
        \text{Layer 1: }& \vphantom{\frac12}f_1(z) = \tanh z; \\
    \text{Layer 2: }& f_2(z) = \operatorname{LeReLU}(\Re z) + \mathrm{i} \operatorname{LeReLU}(\Im z)\\
    \text{Layer 3: }& f_3(\mathbf{z})=\mathbf{z} - \frac{1}{2} \cdot \operatorname{LogSumExp}\left(2 \cdot \Re\mathbf{z}\right).
\end{align*}
Here $\operatorname{LeReLU}$ is a real-valued leaky ReLU function:
\begin{equation*}
\operatorname{LeReLU}(x) =  \begin{cases}
    x, \text{ if } x \geq 0,\\
    -0.01 x \text { otherwise.}
\end{cases}
\end{equation*}
The first two activations are applied elementwise, while the third activation is applied across the whole two-dimensional output vector $\mathbf{z}$ to normalise the probability $\exp(\Re{f_3(z_1)})^2+\exp(\Re{f_3(z_2)})^2$ to unity.  

\subsection{Batch size schedule}\label{app:batch_size_schedule}
In our calculations we are using the following batch size schedule (here $t$ stands for the optimisation iteration index):
\begin{equation*}
    \Ns = \begin{cases}
        10^5,\ 1\leq t \leq 100;\\
        10^6,\ 101\leq  t \leq 200; \\
        10^7,\ 201\leq  t \leq 1000; \\
        10^8,\ t > 1000.
    \end{cases}
\end{equation*}

\subsection{Computing system specifications}\label{app:system_spec}
The calculations are mainly run on NVIDIA RTX A5000 GPU with 24GB of RAM.
The only routine held on a CPU is vectorised sampling of individual binomial distributions; to that end we use 16 Intel(R) Core(TM) i9-11900K CPU cores working at the clock frequency of 3.50GHz.

\section{Ablation studies}\label{app:ablation_studies}
\begin{figure*}
	\centering
	\includegraphics[width=0.75\linewidth]{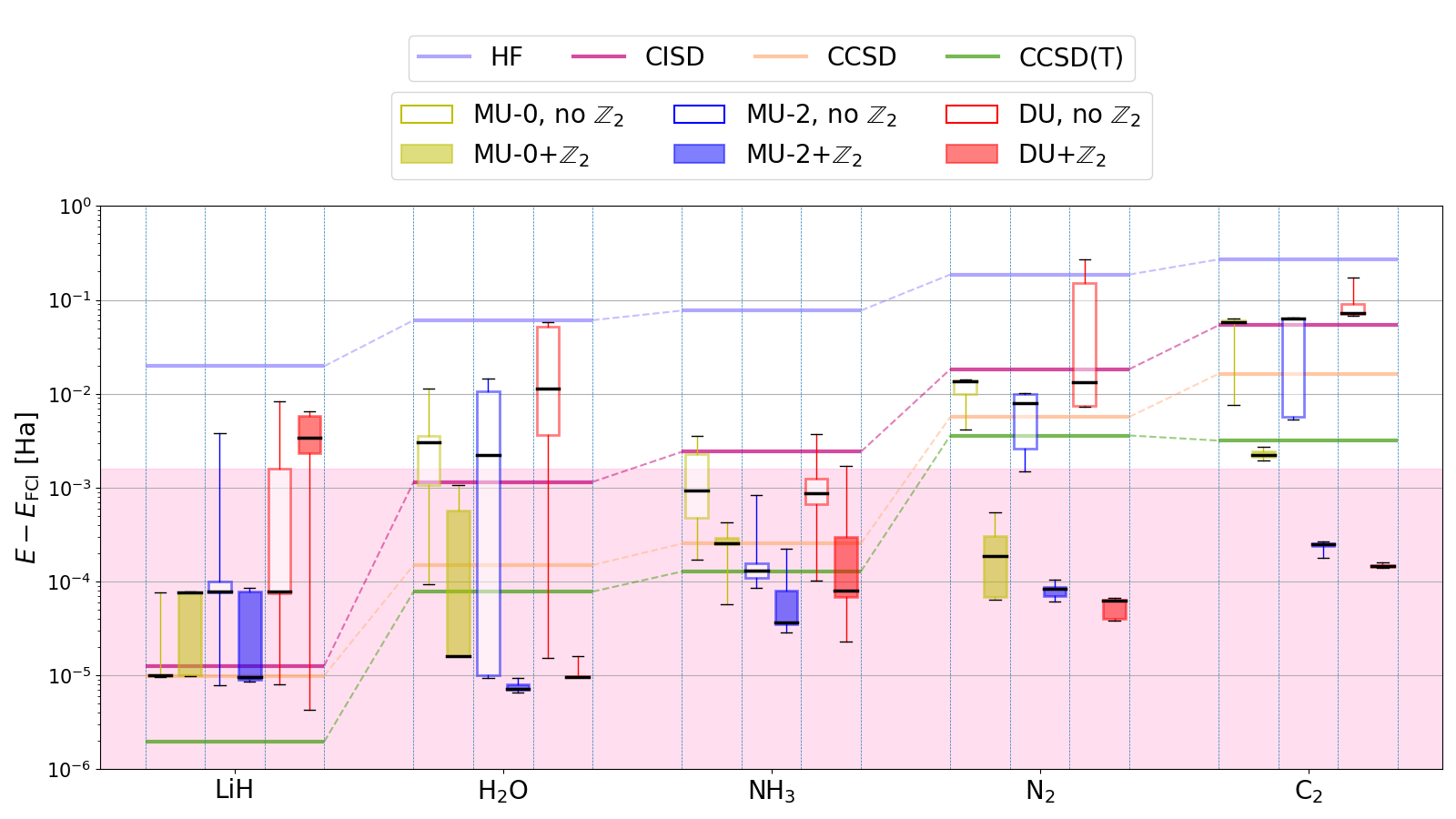}
	\caption{Ablation studies of various local pruning strategies and symmetry sets used for the ANQS-based variational optimisation. Box plots represent five differently initialised ANQS. Chemical accuracy corresponds to the energy difference of $1.6\cdot 10^{-3}$ Ha with respect to FCI. 
 } 
	\label{fig:ablation_studies}
\end{figure*}
In the ablation studies we test how our innovations affected the optimisation performance.
In particular, we (i) demonstrate the effect of taking into account $\mathbb{Z}_2$ symmetries and (ii) compare the pruning strategies described in Sec.~\ref{sec:local_sampling_strategies}.
To that end we run electronic structure calculations for a set of small molecules in the minimal STO-3G basis set. Molecular geometries are taken from Ref.~\cite{carleo_quantum_chemistry}.

The ablation results are presented in Fig.~\ref{fig:ablation_studies}.
We observe that taking $\mathbb{Z}_2$ symmetries into account universally results in better median variational energies.
Particularly striking are the results for highly symmetric molecules such as N$_2$ and C$_2$: the accuracy improvement reaches two orders of magnitude for \MUS{}-2 and \DUS{}  strategies and allows us to achieve the chemical accuracy benchmark of 1.6 mHa. 
As expected, the \MUS{}-2 strategy outperforms \MUS{}.
On the other hand, one can see that the \MUS{}-2 strategy performs on par with \DUS{}.
Hence, in the numerical experiments presented in the main text we employed only \MUS{}-2 and \DUS{} pruning strategies.

\end{document}